 \documentclass{emulateapj}
 \usepackage{apjfonts}
 \usepackage{graphicx}

\shortauthors{Winn et al.\ 2010}
\shorttitle{Oblique Orbit of HAT-P-11b}

\begin{document}

% ------------------------------------------------------------------------
% New commands
%
\def\ltsima{$\; \buildrel < \over \sim \;$}
\def\lsim{\lower.5ex\hbox{\ltsima}}
\def\gtsima{$\; \buildrel > \over \sim \;$}
\def\gsim{\lower.5ex\hbox{\gtsima}}
                                                                                          
% -------------------------------------------------------------------------
%

\bibliographystyle{apj}

\title{ The Oblique Orbit of the Super-Neptune HAT-P-11\lowercase{b} }

\author{
Joshua N.\ Winn\altaffilmark{1},
John Asher Johnson\altaffilmark{2},
Andrew W.\ Howard\altaffilmark{3,4},
Geoffrey W.\ Marcy\altaffilmark{4},
Howard Isaacson\altaffilmark{4},\\
Avi Shporer\altaffilmark{5,6},
G\'asp\'ar \'{A}.\ Bakos\altaffilmark{7},
Joel D.\ Hartman\altaffilmark{7},
Simon Albrecht\altaffilmark{1}
}

\journalinfo{ApJ Letters, in press}
\slugcomment{Received 2010 September 8; accepted 2010 September 27}

\altaffiltext{1}{Department of Physics, and Kavli Institute for
  Astrophysics and Space Research, Massachusetts Institute of
  Technology, Cambridge, MA 02139}

\altaffiltext{2}{Department of Astrophysics, and NASA Exoplanet
  Science Institute, California Institute of Technology, MC~249-17,
  Pasadena, CA 91125}

\altaffiltext{3}{Department of Astronomy, University of California,
  Mail Code 3411, Berkeley, CA 94720}

\altaffiltext{4}{Townes Postdoctoral Fellow, Space Sciences
  Laboratory, University of California, Berkeley, CA 94720}

\altaffiltext{5}{Las Cumbres Observatory Global Telescope Network,
  6740 Cortona Drive, Suite 102, Santa Barbara, CA 93117}

\altaffiltext{6}{Harvard-Smithsonian Center for Astrophysics, 60
 Garden St., Cambridge, MA 02138}

\begin{abstract}

  We find the orbit of the Neptune-sized exoplanet HAT-P-11b to be
  highly inclined relative to the equatorial plane of its host
  star. This conclusion is based on spectroscopic observations of two
  transits, which allowed the Rossiter-McLaughlin effect to be
  detected with an amplitude of 1.5~m~s$^{-1}$. The sky-projected
  obliquity is $103_{-10}^{+26}$~degrees. This is the smallest
  exoplanet for which spin-orbit alignment has been measured. The
  result favors a migration scenario involving few-body interactions
  followed by tidal dissipation. This finding also conforms with the
  pattern that the systems with the weakest tidal interactions have
  the widest spread in obliquities. We predict that the high obliquity
  of HAT-P-11 will be manifest in transit light curves from the {\it
    Kepler} spacecraft: starspot-crossing anomalies will recur at most
  once per stellar rotation period, rather than once per orbital
  period as they would for a well-aligned system.

\end{abstract}


\section{Introduction}
\label{sec:introduction}

The origin of close-in planets is the longest-standing problem in
exoplanetary science (Mayor \& Queloz 1995). Recently, the orbits of
some close-in planets were found to be highly inclined relative to the
equatorial planes of their host stars (see, e.g., H\'ebrard et
al.~2008, Narita et al.~2009, Winn et al.~2009, Triaud et
al.~2010). This evidence supports theories for close-in planets in
which their orbits shrink due to gravitational perturbations from
other bodies followed by tidal dissipation (Matsumura et
al.~2010). The evidence disfavors the other leading theory, in which
the orbits shrink due to gradual interactions with the protoplanetary
gas disk, unless the disks were somehow misaligned with their host
stars (Bate et al.~2010, Lai et al.~2010).

To this point, spin-orbit alignment has been measured only for ``hot
Jupiters,'' with masses ranging from 0.4--20~$M_{\rm Jup}$. We would
like to extend these studies to smaller planets, in order to see
whether they migrate in a similar way as larger planets, and to
understand which factors are associated with orbital misalignment. It
has been claimed, for example, that tilted orbits are more prevalent
for massive planets (Johnson et al.~2009, H\'ebrard et al.~2010), or
for stars with thinner convection zones (Winn et al.~2010, Schlaufman
2010).

Here we present a spin-orbit study of HAT-P-11b, a ``hot Neptune'' of
mass 0.08~$M_{\rm Jup}$ and radius 0.42~$R_{\rm Jup}$ on an eccentric,
4.9-day orbit around a K4 dwarf (Bakos et al.~2010; B10 hereafter). We
observed the Rossiter-McLaughlin (RM) effect (\S~2), and modeled it
(\S~3), finding the orbit and stellar spin to be misaligned (\S~4).

\section{Observations and data reduction}
\label{sec:observations}

We obtained 132 new spectra of HAT-P-11 with the High Resolution
Spectrograph (HIRES; Vogt et al.~1994) on the Keck~I 10m
telescope. Most of the new spectra were gathered on nights when
transits were predicted. On 2009~Aug~2/3 we gathered 7 spectra during
a transit, although fog prevented us from observing before or after
the transit. On 2010~May~26/27 we obtained 32 spectra starting at
around first contact and extending for a few hours beyond the transit.
On 2010~Aug~22/23 we obtained 70 spectra spanning the entire transit
and a few hours beforehand and afterward. The remaining 23 spectra
were obtained sporadically throughout the 2009--2010 observing season.

We used the instrument settings and observing procedures that are
standard for the California Planet Search (Howard et al.~2009). In
particular, we used an iodine gas absorption cell to track the
instrumental response and wavelength scale. The radial velocity (RV)
of each spectrum was measured with respect to an iodine-free template
spectrum, using a descendant of the algorithm of Butler et
al.~(2006). Measurement errors were estimated from the scatter among
the fits to individual spectral segments spanning a few Angstroms.
Table~1 gives all the Keck/HIRES RVs, including re-reductions of the
50 spectra presented by B10.

\section{Analysis}
\label{sec:analysis}

Merging all the RVs into a single analysis requires some care because
the host star is chromospherically active. B10 found a photometric
signal with period 29.2~days and amplitude 3~mmag, which they
attributed to starspots being carried around by stellar rotation. One
would expect a corresponding RV signal at the same period and its
harmonics, with an amplitude of order 0.3\% of the projected stellar
rotation speed ($v\sin i_\star$), or approximately
5~m~s$^{-1}$. Indeed, B10 found evidence for ``stellar jitter'' of
amplitude 5~m~s$^{-1}$, supporting this interpretation.

We investigated this issue by fitting the out-of-transit RVs with a
model consisting of a single Keplerian orbit plus a constant
acceleration\footnote{The evidence for a constant acceleration (linear
  velocity trend), and the implied existence of another orbiting body
  besides HAT-P-11b, were established by B10.}, and seeking evidence
for time-correlated residuals. As seen in Figure~1, the residuals are
strongly correlated on timescales shorter than 5--10~days, as
expected. On longer timescales there are no obvious correlations.

\begin{figure}[ht]
\plotone{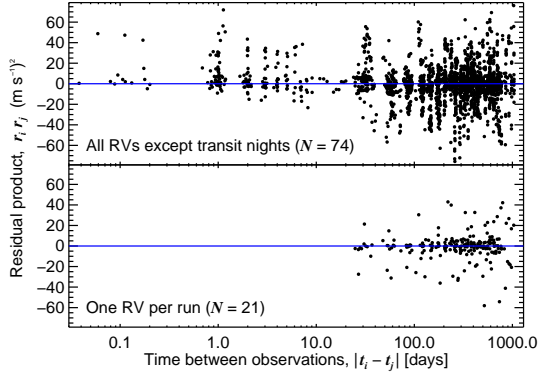}
\caption{{\bf Correlations of the RV residuals.}
{\it Top.}---Products of pairs of residuals, as a function
of the time elapsed between the observations. Significant positive
correlations are seen for $\Delta t \lsim 5$~days.
{\it Bottom.}---Same, but restricting the analysis to only
one data point per observing run, giving a minimum time separation
of 20 days. No significant correlations are seen.
\label{fig:res}}
\end{figure}

While it would be possible to model the RV covariances, we chose the
simpler approach of selecting a subset of RVs that are effectively
independent. Specifically, we chose a single spectrum from each
observing run, resulting in a sample of 21 out-of-transit RVs spaced
apart by a minimum of 20~days.

We turn now to the transit nights. During a single night, one would
expect the rotationally modulated RV signal to act as a nearly
constant offset. Therefore, each transit night was assigned a free
parameter that shifts all the RVs by a common amount; only the
intranight variations were deemed significant. With this approach the
data from 2009~Aug~2/3 were rendered useless, because no data were
gathered outside of the transit. We omitted those data from further
consideration.

Our model for the RVs was the sum of the Keplerian orbital motion, a
constant acceleration, the RM effect, and the offsets described
above. The fitting statistic was
\begin{eqnarray}
\chi^2 & = & \sum_{i=1}^{123}
\left[ \frac{ {\rm RV}_i({\rm obs}) - {\rm RV}_i({\rm calc}) }
  {\sigma_i} \right]^2 + \;\;\;\nonumber \\
&  &   \left(\frac{{\rm BJD}_c - 2\,454\,605.89132}{0.00032}\right)^2 + 
     \left(\frac{P_{\rm days} - 4.8878162}{0.0000071}\right)^2 + \;\;\;\nonumber \\
&  &   \left(\frac{T_{\rm days} - 0.0957}{0.0012}\right)^2 + 
     \left(\frac{\tau_{\rm days} - 0.0051}{0.0013}\right)^2 + \;\;\;\nonumber \\
&  &   \left(\frac{R_p/R_\star - 0.0576}{0.0009}\right)^2 + 
     \left(\frac{R_\star/R_\odot - 0.752}{0.021}\right)^2 + \;\;\;\nonumber \\
&  &   \left(\frac{v\sin i_\star - 1.5~{\rm km~s}^{-1}}{1.5~{\rm km~s}^{-1}}\right)^2,
\end{eqnarray}
where the first term is the usual sum of squared residuals, and the
other terms represent {\it a priori} constraints on parameters that
were determined more precisely from the larger body of data analyzed
by B10. In this expression $P_{\rm days}$ is the orbital period in
days, BJD$_c$ is a particular time of inferior conjunction; $T_{\rm
  days}$ is the time between first and fourth contact; $\tau_{\rm
  days}$ is the time between first and second contact; $R_p$ and
$R_\star$ are the radii of the planet and star; and $v\sin i_\star$ is
the star's sky-projected rotation speed.

Each of the 21 orbital RVs was assigned an error bar $\sigma_i$ equal
to the quadrature sum of the measurement error and a ``jitter'' of
5.5~m~s$^{-1}$, the value giving $\chi^2 = N_{\rm dof}$ when the
orbital RVs were fitted alone. For the transit-night RVs, the jitter
was fixed by the requirement $\chi^2 = N_{\rm dof}$ when fitting the
data from that night along with the orbital RVs. The results were 1.8
and 1.5~m~s$^{-1}$ for 2010~May~26/27 and 2010~Aug~22/23,
respectively.  The relative smallness of these values corroborates our
assumption that the activity-induced RV variations occur mainly on
longer timescales. A similar contrast between intranight and
internight jitter was observed previously for HD~189733, another
active K star (Winn et al.~2006).

We modeled the RM effect with the technique described by Winn et
al.~(2005), finding in this case that a sufficiently accurate
description for the RV shift is the product of the loss of light and
the RV of the portion of the stellar photosphere beneath the planet.
We neglected differential rotation, and took the stellar
limb-darkening law to be linear with a coefficient of 0.79
(Claret~2004).

Parameter optimization and error estimation were achieved with a
Markov Chain Monte Carlo algorithm, using Gibbs sampling and
Metropolis-Hastings stepping. Table~2 gives the results for each
parameter, based on the 15.85\%, 50\%, and 84.15\% confidence levels
of the marginalized {\it a posteriori} distributions. Figure~2 shows
the RV data: the left panel shows the orbital RVs; and the right panel
shows the transit-night RVs after subtracting the calculated variation
due to orbital motion, thereby isolating the ``anomalous RV'' due to
the RM effect. Figure~3 shows the {\it a posteriori} distributions for
the key parameters $v\sin i_\star$ and $\lambda$ (the sky position
angle from the stellar north pole to the orbital north pole).

\begin{figure*}[ht]
\begin{center}
\leavevmode
\hbox{
\epsfxsize=7.5in
\epsffile{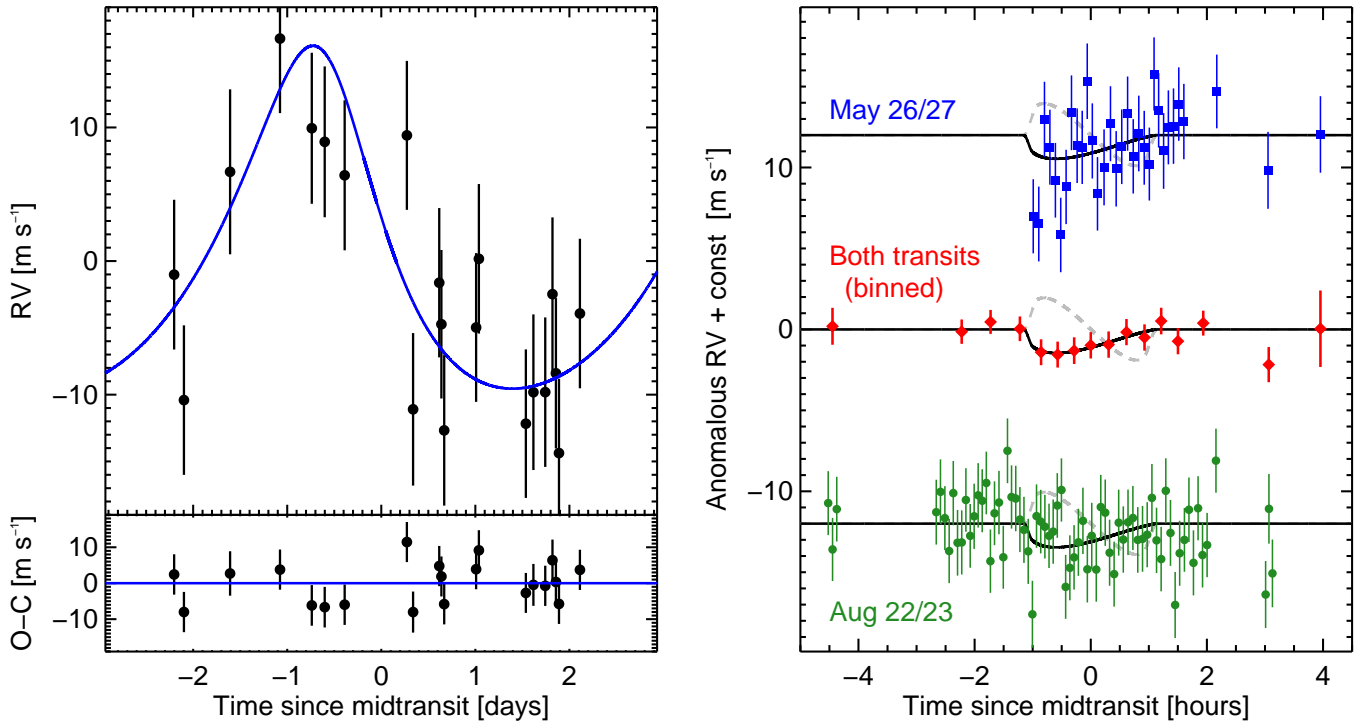}}
\end{center}
\vspace{-0.2in}
%\begin{figure}[ht]
%\plotone{rv.eps}
\caption{{\bf Radial velocities and the RM effect.}  {\it
    Left.}---Spectroscopic orbit of HAT-P-11, based on the subset of
  21 RVs analyzed here.  {\it Right.}---Anomalous RV of HAT-P-11
  spanning two transits (top and bottom), and the time-binned
  combination (middle; binned~$\times$7 with a maximum bin size of
  0.5~hr). The orbital contribution to the RV has been subtracted.
  The solid line shows the best-fitting model of the RM effect.  The
  dashed curve shows the best-fitting ``well-aligned'' model
  ($\lambda=0$, $v\sin i_\star = 1.3$~km~s$^{-1}$), which is ruled out
  with $\Delta\chi^2 = 52.4$.
  \label{fig:rv}}
\end{figure*}
%\end{figure}

\begin{figure*}[ht]
\begin{center}
\leavevmode
\hbox{
\epsfxsize=7.5in
\epsffile{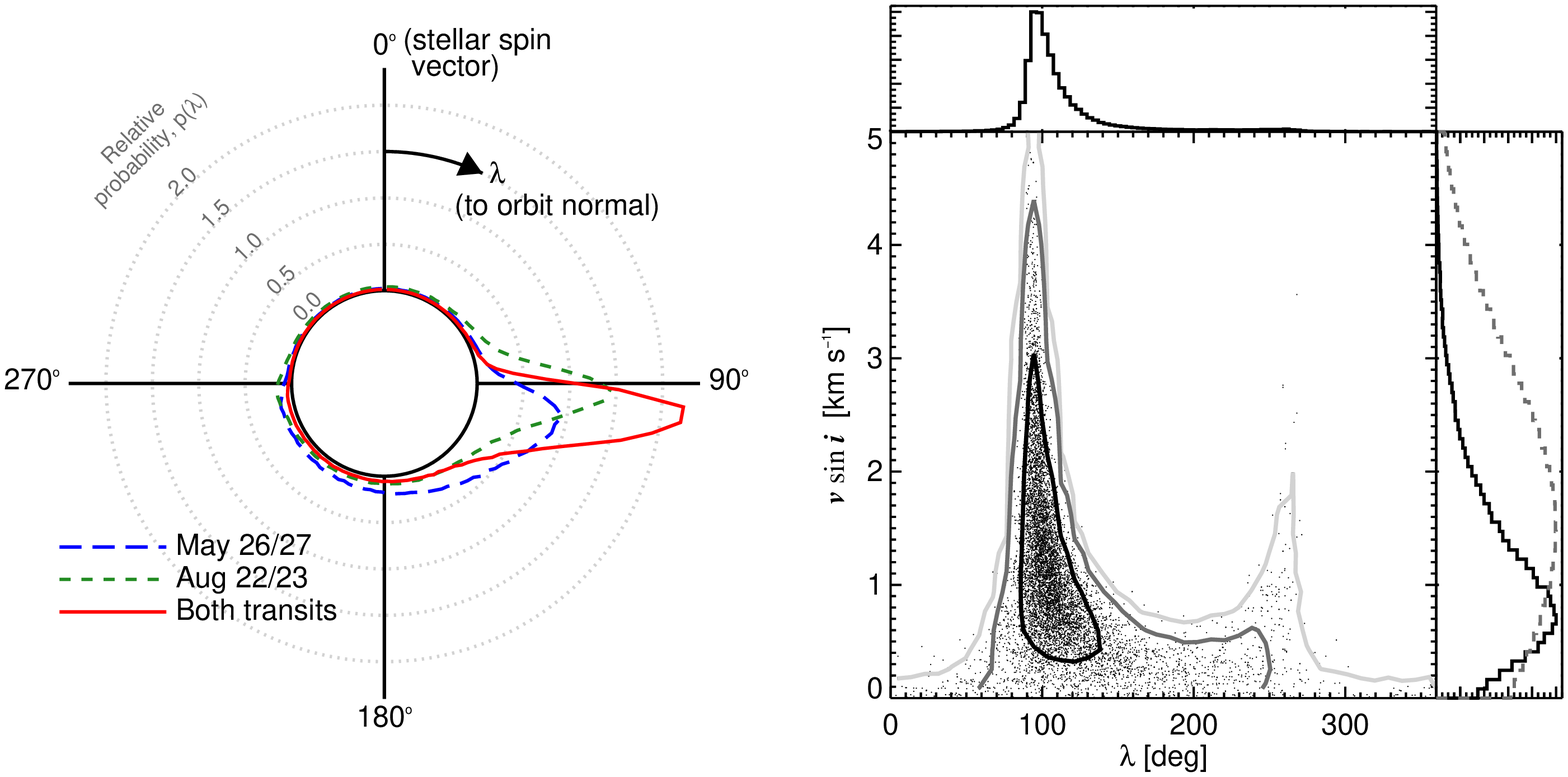}}
\end{center}
\vspace{-0.2in}
%\begin{figure}[ht]
%\plotone{mcmc.eps}
\caption{{\bf Results for the RM parameters.}
{\it Left.}---In this polar plot, the angular coordinate is $\lambda$
and the radial coordinate is $p(\lambda)$, the marginalized
posterior probability distribution. Results are shown
for analyses based only on individual transits, as well
as for the entire dataset.
{\it Right.}---Joint constraints on $\lambda$ and $v\sin i_\star$.
The contours represent 68\%, 95\%, and 99.73\%
confidence limits. The
marginalized posterior probability distributions
are shown on the sides of the contour plot. The dashed histogram
shows the {\it a priori} constraint on $v\sin i_\star$.
\label{fig:mcmc}}
\end{figure*}
%\end{figure}

The most important result is $\lambda=103_{-10}^{+26}$~degrees,
indicating a major misalignment between the stellar rotation axis and
the orbit normal. Qualitatively this follows from the observation that
the RM effect was observed to be a blueshift throughout the transit,
as opposed to the ``red-then-blue'' pattern of a well-aligned system.

\section{ Discussion }
\label{sec:discussion}

Because the signal has an amplitude of only 1.5~m~s$^{-1}$, smaller
than any other RM signal yet reported, it is important to test the
robustness of the results. First, we tried analyzing each transit
individually, rather than combining the data from both transits. As
shown in Figure~3, the results are in good agreement.\footnote{In fact
  there are three mutually reinforcing datasets: Hirano et al.\ have
  submitted a paper reporting $\lambda = 103_{-19}^{+23}$~degrees,
  based on independent observations of HAT-P-11b (T.\ Hirano and N.\
  Narita, private communication).} Second, we repeated the analysis
without the {\it a priori} constraint on $v\sin i_\star$, the other
parameter of greatest relevance to the RM effect. The results were
$v\sin i_\star = 1.13_{-0.70}^{+2.44}$~km~s$^{-1}$ and
$\lambda=100_{-9}^{+28}$~degrees. Third, we checked for any
correlations between the RM signal and the strength of Ca~II H and K
emission, or the shape parameters of the instrumental line spread
function. Significant correlations would have raised suspicion of
systematic errors, but none were found.

In addition, if the 29.2~day periodicity detected by B10 is indeed the
rotation period, then a powerful test for spin-orbit misalignment is
available. If the system were well-aligned we would have $\lambda = 0$
and $i_\star \approx 90^\circ$. This would imply
\begin{equation}
v\sin i_\star = v = \frac{2\pi R_\star}{P_{\rm rot}} = 1.30~{\rm km~s}^{-1},
\end{equation}
where we have used $R_\star = 0.752$~$R_\odot$ and $P_{\rm rot} =
29.2$~days (B10). When refitting the data with these constraints, the
minimum $\chi^2$ rises from 111.9 to 164.3 ($\Delta\chi^2 = 52.4$),
with 114 degrees of freedom. Thus the well-aligned model is ruled out
with 7.2$\sigma$ confidence: either $\lambda$ is large, or else
$i_\star$ must be far from $90^\circ$ to be compatible with the low
amplitude of the RM signal. The best-fitting well-aligned model is
illustrated with a gray dashed curve in Figure~3.

HAT-P-11b is the first ``hot Neptune'' for which the RM effect has
been measured. Our results suggest that tilted orbits are common for
hot Neptunes, just as has been found for hot Jupiters. The same
migration mechanisms that are invoked to explain the larger planets
with tilted orbits---gravitational scattering by planets, or the
three-body Kozai effect---may also have operated in this case.  It
should be noted that the spin-orbit results are not the only evidence
for a perturbative origin for many close-in planets.  Further evidence
comes from their occasionally high orbital eccentricities, the
clustering of their orbital distances near the value expected from
tidal circularization, and their tendency to lack companions with
periods between 10--100~days (Matsumura et al.~2010).

Since HAT-P-11b is the lowest-mass planet yet probed by RM
measurements, and it is misaligned, our findings are at odds with the
hypothesis that misalignments occur mainly for the most massive
planets (Johnson et al.~2009). They do, however, support the
correlation between large obliquity and orbital eccentricity (Johnson
et al.~2009, H\'ebrard et al.~2010), as the orbit of HAT-P-11b has a
significant eccentricity.

Another emerging trend is that misalignments occur mainly for stars
with high effective temperatures or large masses ($T_{\rm eff}>6250$~K
or $M_\star \gsim 1.2~M_\odot$). Winn et al.~(2010) speculated that
this is due to tidal interactions: cool stars realign with the orbits,
but hot stars cannot realign because tidal dissipation is weaker in
their thinner outer convection zones. The HAT-P-11 system is an
important test case because the star is cool and low-mass, and yet
tidal interactions are weak due to the planet's relatively small size
and long period. If stellar temperature or mass are the determinants
then one would expect HAT-P-11 to be well-aligned like other cool
stars. But if tides are the underlying factor, then HAT-P-11 would be
misaligned, as we have observed. Specifically, with reference to
Eqn.~(2) of Winn et al.~(2010), HAT-P-11 experiences even weaker tides
than WASP-8, a cool star already known to have a high
obliquity.\footnote{Matsumura et al.~(2010) have also argued that
  tidal dissipation in the HAT-P-11 system would be too weak to
  realign the star with the orbit, despite the star's thick outer
  convective zone.} Thus, HAT-P-11 is a telling exception to the rule
that hot stars have high obliquities: it implicates tidal evolution as
the reason for low obliquities among cool stars with more massive
planets in tighter orbits.

By good fortune, HAT-P-11 is in the field of view of the {\it Kepler}
spacecraft (Borucki et al.~2010). The precise photometric time series
will allow the candidate 29.2-day rotation period to be
checked. Asteroseismological studies may reveal the stellar mean
density, age, inclination, and other parameters (Christensen-Dalsgaard
et al.~2010). Furthermore we predict that {\it Kepler} will see a
pattern of anomalies in the transit light curves that will betray the
system's spin-orbit misalignment. As usual for a spotted star, there
will be a ``bump'' or ``rebrightening'' in the transit light curve
whenever the planet occults a starspot (see, e.g., Rabus et
al.~2009). For a well-aligned star, the bumps recur in successive
transits for as long as the spot is on the visible hemisphere. There
is a steady advance in phase of the bumps due to the star's rotation
between transits. However, for a star like HAT-P-11 with
$\lambda\approx 90^\circ$, the events will {\it not} recur in this
manner, because the star's rotation moves the spot away from the
transit chord. A spot must complete a full rotation before returning
to the transit chord, and even then, the planet will miss it unless it
has also completed an integral number of orbits.

For HAT-P-11, it happens that $P_{\rm rot}/P_{\rm orb} \approx 6$. If
the star were well-aligned and had one spot initially on the transit
chord, then we would typically see an alternation between 2--3 light
curves with bumps, and 2--3 without bumps (when the spot is on the far
side). But because of the misalignment, spot anomalies will only recur
every 29.2~days, after the star has rotated once and the planet has
completed 6 orbits. Complications may arise due to differential
rotation, as well as the multiplicity and evolution of spot
patterns. Nevertheless, this phenomenon should allow for an
independent test of spin-orbit misalignment for HAT-P-11 as well as
other spotted stars.

\acknowledgements We thank Norio Narita and Teruyuki Hirano for
sharing their results prior to publication, and Dan Fabrycky and Scott
Gaudi for helpful discussions. We acknowledge the support from the MIT
Class of 1942, NASA grants NNX09AD36G and NNX08AF23G, and NSF grant
AST-0702843.

The data presented herein were obtained at the W.~M.~Keck Observatory,
which is operated as a scientific partnership among the California
Institute of Technology, the University of California, and NASA, and
was made possible by the generous financial support of the W.~M.~Keck
Foundation. We extend special thanks to those of Hawaiian ancestry on
whose sacred mountain of Mauna Kea we are privileged to be
guests.\\

\clearpage

\begin{deluxetable}{lcc}

\tabletypesize{\scriptsize}
\tablecaption{Relative Radial Velocity Measurements of HAT-P-11\label{tbl:rv}}
\tablewidth{0pt}

\tablehead{
\colhead{BJD$_{\rm UTC}$} &
\colhead{RV [m~s$^{-1}$]} &
\colhead{Error [m~s$^{-1}$]}
}

\startdata
  $  2454335.89332$  &  $     -1.33$  &  $   1.01$  \\
  $  2454335.89998$  &  $     -2.66$  &  $   1.03$  \\
  $  2454336.74876$  &  $     -1.30$  &  $   0.90$  \\
  $  2454336.86163$  &  $     -4.74$  &  $   1.02$  \\
  $  2454336.94961$  &  $     -7.56$  &  $   0.94$
\enddata

\tablecomments{The RV was measured relative to an arbitrary template
  spectrum; only the differences are significant. The uncertainty
  given in Column 3 is the internal error only and does not account
  for any possible ``stellar jitter.'' (We intend for this table to be
  available in its entirety in a machine-readable form in the online
  journal. A portion is shown here for guidance regarding its form and
  content.)}

\end{deluxetable}

\begin{deluxetable}{lc}
 
\tabletypesize{\normalsize}
\tablecaption{Model Parameters\label{tbl:params}}
\tablewidth{0pt}
 
\tablehead{
\colhead{Parameter} &
\colhead{Value}
}

\startdata \hline
{\it Parameters controlled mainly by priors} \\
\hline
Orbital period, $P$~[days]                         & $4.88781501 \pm 0.0000068$ \\
Midtransit time~[BJD$_{\rm UTC}$]                    & $2,454,605.89130 \pm 0.00032$ \\
Planet-to-star radius ratio, $R_p/R_\star$           & $0.0576\pm 0.00090$ \\ 
Orbital inclination, $i$~[deg]                     & $89.17_{-0.60}^{+0.46}$     \\ 
Fractional stellar radius, $R_\star/a$              & $0.0673\pm 0.0018$    \\ 
\hline
{\it Parameters controlled mainly by RV data} \\
\hline
Velocity semiamplitude, $K$~[m~s$^{-1}$]            & $12.9 \pm 1.4$ \\
$e\cos\omega$                                      & $0.261\pm 0.082$ \\
$e\sin\omega$                                      & $0.085\pm 0.043$ \\
RV offset, 2010~May~26/27 [m~s$^{-1}$]            & $-19.8 \pm 3.9$ \\ 
RV offset, 2010~Aug~22/23 [m~s$^{-1}$]            & $-17.5 \pm 4.2$ \\ 
RV offset, all other data [m~s$^{-1}$]            & $-13.0 \pm 2.2$ \\ 
RV trend, $\dot{\gamma}$ [m~s$^{-1}$~day$^{-1}$]  & $0.0185\pm 0.0036$ \\
Projected stellar rotation rate, $v \sin i_\star$~[km~s$^{-1}$]   &  $1.00_{-0.56}^{+0.95}$ \\
Projected spin-orbit angle, $\lambda$~[deg]                     &  $103_{-10}^{+26}$
\enddata

\end{deluxetable}

\end{document}